\documentclass[a4paper,11pt]{article}
\usepackage[dvips]{graphicx}
\usepackage{cite}

\begin{document}

\title{Nanophotonic enhancement of the F\"{o}rster resonance energy transfer rate on single DNA molecules}

\author{Petru Ghenuche, Juan de Torres, Satish Babu Moparthi, Victor Grigoriev, and J\'{e}r\^{o}me Wenger$^{*}$}

\date{}

\maketitle

CNRS, Aix-Marseille Universit\'e, Centrale Marseille, Institut Fresnel, UMR 7249, 13013 Marseille, France

$^*$ Corresponding author: jerome.wenger@fresnel.fr

\vspace{1.5cm}


\begin{abstract}
Nanophotonics achieves accurate control over the luminescence properties of a single quantum emitter by tailoring the light-matter interaction at the nanoscale and modifying the local density of optical states (LDOS). This paradigm could also benefit to F\"{o}rster resonance energy transfer (FRET) by enhancing the near-field electromagnetic interaction between two fluorescent emitters. Despite the wide applications of FRET in nanosciences, using nanophotonics to enhance FRET remains a debated and complex challenge. Here, we demonstrate enhanced energy transfer within single donor-acceptor fluorophore pairs confined in gold nanoapertures. Experiments monitoring both the donor and the acceptor emission photodynamics at the single molecule level clearly establish a linear dependence of the FRET rate on the LDOS in nanoapertures. These findings are applied to enhance the FRET rate in nanoapertures up to six times, demonstrating that nanophotonics can be used to intensify the near-field energy transfer and improve the biophotonic applications of FRET.
\end{abstract}

Nanoscale energy transfer between molecules is a core phenomenon in photosynthesis \cite{Kuhlbrandt91,Grondelle94,Niek13} and an enabling technology for photovoltaics \cite{Farrell09,Shankar09}, organic lighting sources \cite{Baldo00} or biosensing \cite{Medintz03}. When the distance between the excited donor D to the ground-state acceptor molecule A is on the range of 2-20~nm, the energy transfer is adequately described by the F\"{o}rster resonance energy transfer (FRET) formalism which accounts for near-field nonradiative dipole-dipole interaction \cite{Forster48}. As the FRET efficiency goes down with the inverse sixth power of the D-A distance, FRET provides accurate information with sub-nanometer resolution on the spatial relationship between two fluorophore-labeled sites in biological structures \cite{Roy08}. FRET has thus become one of the most popular tools in single molecule spectroscopy, and is largely used to study conformational changes in macromolecules as well as molecular interaction dynamics between proteins, DNA, RNA and peptide molecules \cite{Weiss00,Schuler08}.

The interaction between molecules and light at the nanoscale is at the heart of FRET. Therefore using nanophotonics to control light at the nanoscale is appealing to enhance the near-field dipole-dipole energy transfer. Since the pioneering works of Purcell and Drexhage \cite{Purcell46,Drexhage70}, it is well established that the luminescence properties of a single quantum emitter can be engineered by its photonic environment through the local density of optical states (LDOS) \cite{Novotnybook,Barnes98}. Successful demonstrations of this concept include cavity quantum electrodynamics \cite{Haroche}, photonic band gap materials \cite{Yablo,Lodahl04}, and more recently plasmonic antennas \cite{NovotnyRev11,Moerner09}. However, when the discussion turns to the interaction of two emitters with their environment, the LDOS influence on the F\"{o}rster energy transfer remains an open debate. Since FRET directly competes with the donor direct emission and the donor nonradiative energy losses to its environment, it is unclear whether F\"{o}rster transfer can be enhanced when the LDOS is tuned to increase the donor emission. Pioneering works on ensemble measurements in microcavities suggested that the FRET rate depends \textit{linearly} on the donor emission rate and the LDOS \cite{Hopmeier99,Andrew00,Finlayson01,Nakamura05}. While several theoretical works support this finding \cite{Dung,Colas,Govorov07,Carminati,Enderlein}, it was also suggested that the FRET rate was \textit{quadratically} dependent on the LDOS \cite{Nakamura06}, or even \textit{independent} of the LDOS \cite{Polman05}. A recent experimental study on molecules near a planar mirror concludes that the energy transfer rate is \textit{independent} of the LDOS \cite{Blum12}. In this case, the FRET efficiency and characteristic distance are reduced when the LDOS is enhanced, which comes in apparent contradiction with qualitative photonic effects on FRET published earlier \cite{Lakowicz07,Fore07,Kolaric07,Hohenester08,Yang08,Bradley08,Bradley11,Bradley12}. Moreover, most studies consider only the donor emission photodynamics and measure ensembles of molecules.

\begin{figure}
\begin{center}
\includegraphics{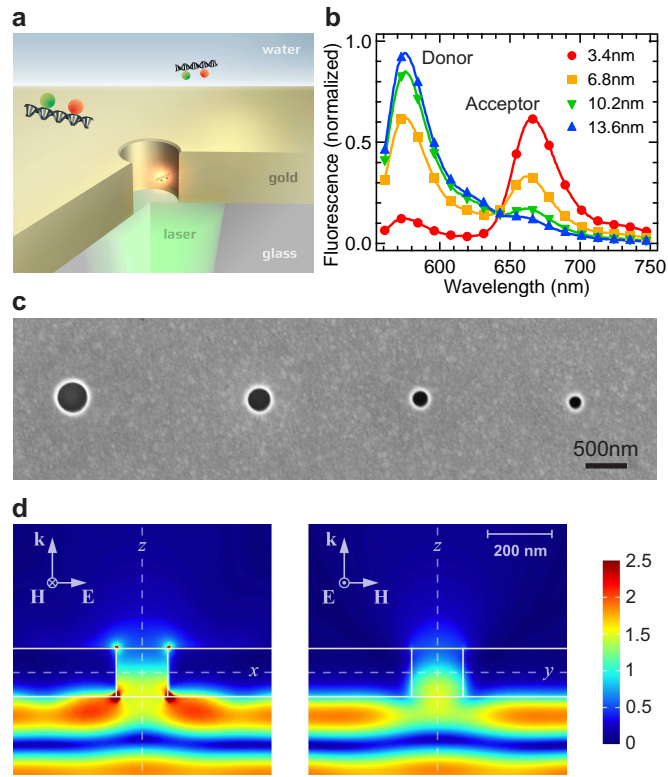}
\caption{\textbf{Single molecule FRET in a gold nanoaperture.} (a) Schematic of a donor-acceptor pair on a DNA molecule diffusing inside and around a single nanoaperture milled in a gold film on a glass coverslip. (b) Normalized fluorescence emission spectra for different donor-acceptor distances in water solution. Emission of the Atto550 donor at 580~nm is recovered while emission of the Atto647N acceptor at 670~nm vanishes as the donor-acceptor distance is increased from 3.4 to 13.6~nm. (c) Scanning electron microscopy image of nanoapertures of diameters from 380 to 160~nm. (d) Finite-element computation of excitation electric field amplitude enhancement for two orthogonal planes on a gold nanoaperture of 160~nm diameter. The incoming light at a wavelength of 550~nm is polarized along the x axis.} \label{FigIntro}
\end{center}
\end{figure}

Here, we explore the LDOS influence on the FRET process using precisely defined plasmonic nanoapertures to control the LDOS and single donor-acceptor fluorophore pairs on double-stranded DNA linkers to provide a wide range of FRET rates and efficiencies (Fig.~\ref{FigIntro}a). Our experiments are conducted at the single molecule level and monitor both the donor and the acceptor emission photodynamics. Two separate measurements of the FRET rate $\Gamma_{FRET}$ and efficiency $E_{FRET}$ are obtained based either on the donor lifetime reduction in the presence of the acceptor, or the acceptor fluorescence increase in presence of the donor. The FRET efficiency is defined as the probability of energy transfer over all donor transition events: $E_{FRET}=\Gamma_{FRET}/\Gamma_{DA}=\Gamma_{FRET}/(\Gamma_{FRET}+\Gamma_{D})$, where $\Gamma_{D}$ is the total decay rate of the isolated donor (without acceptor) and $\Gamma_{DA}=\Gamma_{D}+\Gamma_{FRET}$ is the total decay rate of the donor in the presence of the acceptor. The donor-only decay rate $\Gamma_D$ is proportional to the LDOS at the donor emission wavelength and is tuned by the photonic environment \cite{Novotnybook,Barnes98}.

To provide a wide range of LDOS, we use circular nanoapertures with diameters ranging from 160 to 380~nm milled in a 150~nm thick gold film on a glass coverslip by focused ion beam (Fig.~\ref{FigIntro}c,d). Molecules randomly diffusing inside the apertures experience an average fluorescence lifetime reduced by a factor up to three fold. While the LDOS is expected to vary spatially inside the nanoaperture depending on the position and orientation of the emitter respective to the metal, the net fluorescence lifetime reduction observed on the decay traces averaged for all emitter positions and orientations inside the aperture demonstrates an increase of the spatially-averaged LDOS inside nanoapertures \cite{Rigneault05,Wenger08,LutzACS13}. Nanoapertures thus realize a reproducible platform to tune the spatially-averaged LDOS. They also have a broad spectral response so that the same structure can enhance both the donor and the acceptor emission (see Supplementary Information Fig.~7 and 8).

To provide a wide range of FRET rates and efficiencies, we synthesize double stranded DNA molecules with increasing distances between the Atto550 donor and the Atto647N acceptor from 10 to 40 base pairs (3.4 to 13.6~nm). The double stranded DNA forms a rigid linker enabling accurate positioning the donor and acceptor with sub-nanometer resolution \cite{Deniz99}. The spectral overlap between Atto550 and Atto647N ensures F\"{o}rster transfer with a characteristic distance $R_0$ of 6.5~nm in pure water medium, that is confirmed by ensemble fluorescence spectroscopy (Fig.~\ref{FigIntro}b). Hence by varying either the donor-acceptor distance or the nanoaperture diameter, we have two independent parameters to tune the FRET rate and the LDOS, and monitor the effect of the LDOS on the FRET rate.

\begin{figure}
\begin{center}
\includegraphics{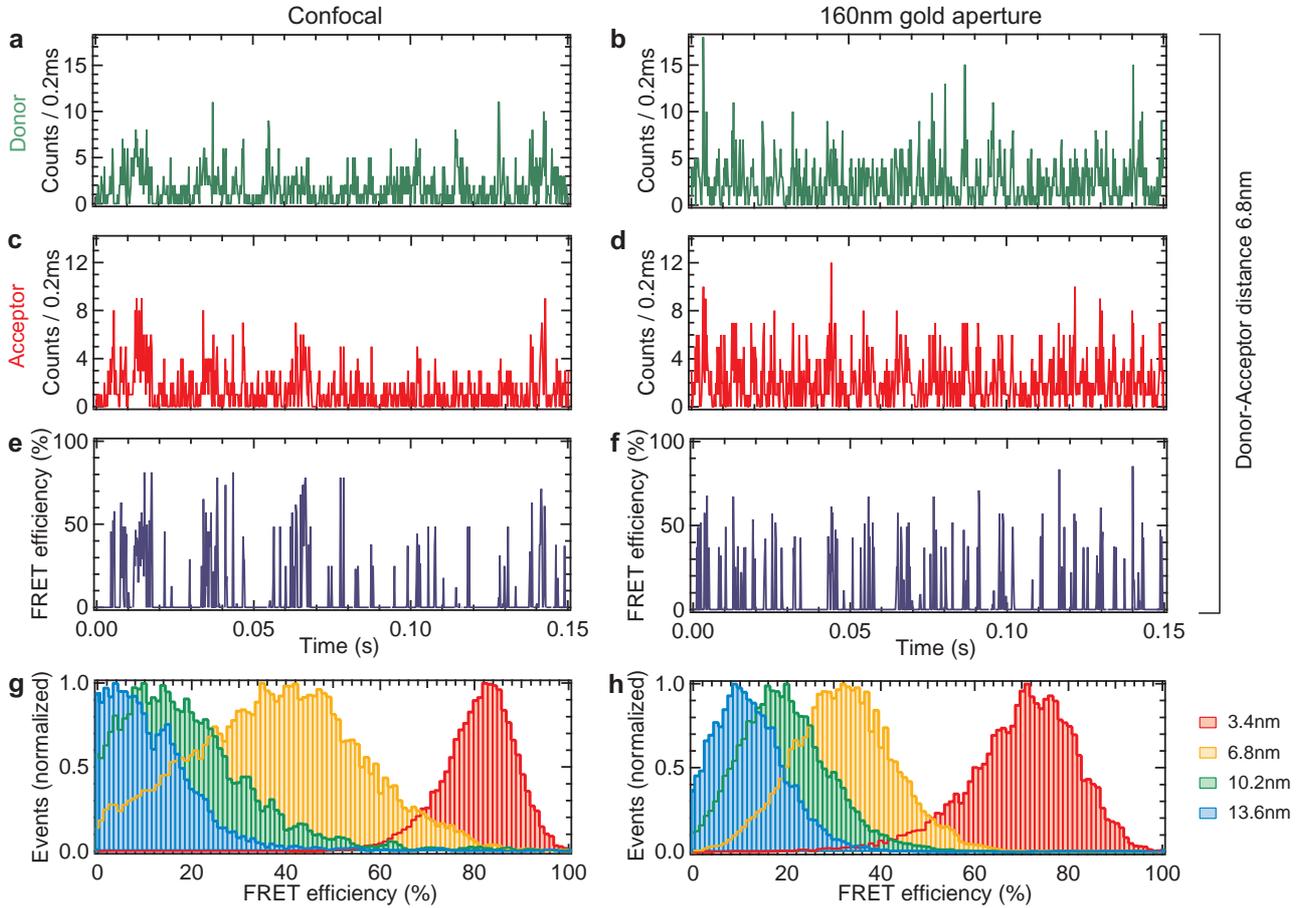}
\caption{\textbf{FRET burst analysis of single molecule diffusion events in a  gold nanoaperture}. (a-f) Example of time traces for donor-acceptor pairs separated by 6.8~nm (20 base pairs) diffusing in solution and probed by the diffraction-limited confocal microscope (a,c,e) and in a gold nanoaperture of 160~nm diameter (b,d,f). The concentration is set so as to have less than one molecule in the detection volume in each case (confocal 1~nM, nanoaperture 250~nM). For each fluorescence burst exceeding the detection threshold in the donor detection channel (a,b) or the acceptor channel (c,d), a FRET efficiency is calculated (e,f). The binning time is 0.2~ms and the total trace duration is 200~s. (g,h) FRET efficiency histograms extracted from fluorescence burst analysis for different donor-acceptor separations in confocal (g) and in the 160~nm nanoaperture (h).} \label{FigBursts}
\end{center}
\end{figure}

For direct observation of single molecule diffusion events, the nanoaperture sample is covered by a water solution containing the diluted DNA FRET pairs. Every time a fluorescent molecule crosses the detection volume a fluorescence burst is generated, which results in the typical time traces displayed on Fig.~\ref{FigBursts}a-d. The burst duration is set by the translational diffusion time and typically amounts to 460~$\mu$s for the confocal reference and 135~$\mu$s for the nanoaperture of 160~nm diameter. Comparing the bursts intensity in confocal to the nanoaperture, a fluorescence enhancement of about 2 times is found for the donor and the acceptor detection channels (see Supplementary Information Fig.~6 and 7 for fluorescence correlation spectroscopy analysis and fluorescence enhancement factors). For each fluorescence photon burst, a FRET efficiency $E_{FRET}$ can be computed (see Methods) and collected in a histogram (Fig.~\ref{FigBursts}e,f). Figures~\ref{FigBursts}g,h summarize the FRET efficiency histograms for increasing donor-acceptor distances in confocal and in the 160~nm nanoaperture. Except for the 3.4~nm D-A distance, the histogram width is reduced by about 50\% inside the nanoaperture as compared to confocal reference. This is primarily a consequence of the higher photon counts per emitter (fluorescence enhancement) in the nanoaperture. The average FRET efficiencies also appear marginally affected by the nanoaperture, with a different influence on the FRET efficiency depending on the D-A distance: while the FRET efficiency for the shortest D-A distance (3.4~nm) is reduced by the nanoaperture from 83 to 76\%, the FRET efficiency for the longest D-A distance (13.6~nm) is increased from 4 to 10\%.

\begin{figure}
\begin{center}
\includegraphics{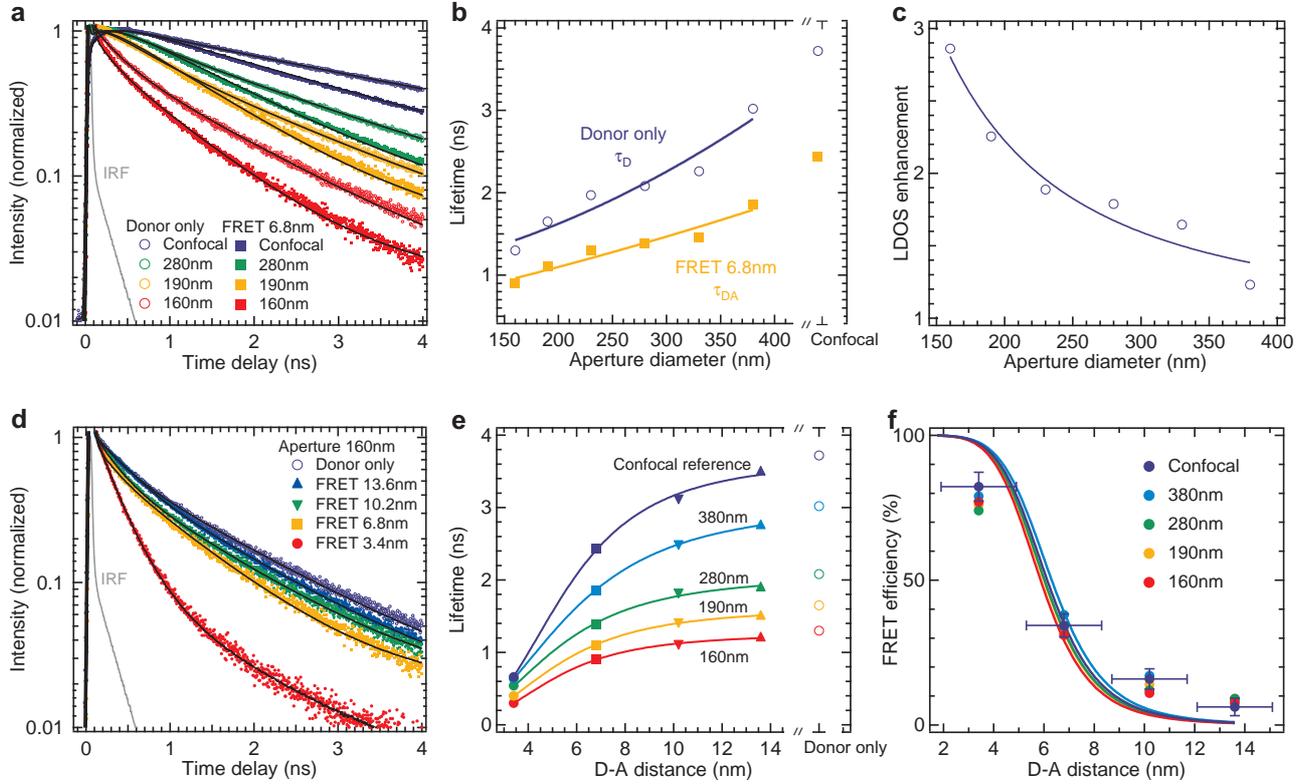}
\caption{\textbf{Donor lifetime reduction in nanoapertures}. (a) Normalized donor decay traces when no acceptor is present (empty circles) or when the acceptor is separated by 6.8~nm (filled squares). The black lines are single-exponential numerical fits convoluted by the instrument response function (IRF). From top to bottom, the different colors are associated with the confocal case or with nanoapertures of decreasing diameters from 280 to 160~nm. In each case, the presence of the acceptor induces a significant acceleration of the donor emission dynamics. (b) Donor fluorescence lifetime as function of the nanoaperture diameter, obtained from the data shown in (a). (c) LDOS enhancement as function of the nanoaperture diameter, obtained from the donor-only lifetime reduction as compared to the confocal reference. (d) Normalized donor decay traces in a 160~nm nanoaperture as the acceptor separation is reduced. (e) Donor fluorescence lifetime as function of the donor-acceptor separation for different nanoaperture diameters and for the confocal reference. For each donor-acceptor distance, a clear reduction of the donor fluorescence lifetime is observed as the nanoaperture diameter is reduced. (f) FRET efficiency deduced from the donor lifetime reduction in the presence of the acceptor, as function of the donor-acceptor distance $R$. The lines are numerical fits assuming a $1/R^6$ dependence. Vertical error bars indicate one standard deviation, horizontal errors bars assume a 1.5~nm distance uncertainty which we relate to the dye-DNA linker flexibility.} \label{FigTCSPC}
\end{center}
\end{figure}

To quantify the acceleration of the fluorescence photodynamics and the LDOS enhancement in nanoapertures, we record the donor fluorescence decay traces by time-correlated single photon counting. Figures~\ref{FigTCSPC}a-c report the influence of the nanoaperture diameter (see also Supplementary Information Fig.~8). For the isolated donor, the fluorescence lifetime $\tau_D$ decreases from 3.7 to 1.3~ns as the nanoaperture diameter is reduced to 160~nm. This data is equivalent to an increase in the donor-only total decay rate $\Gamma_D=1/\tau_D$ and the LDOS by a factor 2.9 (Fig.~\ref{FigTCSPC}c). In the presence of the acceptor at a 6.8~nm separation, the donor emission dynamics are further accelerated by the additional decay channel brought by the acceptor: $\Gamma_{DA}=\Gamma_{D}+\Gamma_{FRET}$ and the donor fluorescence lifetime $\tau_{DA}=1/\Gamma_{DA}$ is further reduced (Fig.~\ref{FigTCSPC}b). We monitor a lifetime reduction as function of the aperture diameter similar to the case of the donor only, from 2.5 to 0.9~ns in the presence of the acceptor. Figures~\ref{FigTCSPC}d-f then describe the influence of the donor-acceptor separation for a given nanoaperture diameter. A clear reduction of the donor emission lifetime (donor quenching) is observed as the D-A distance is reduced, with a similar trend observed for all nanoaperture diameters. The decay rates in the presence and absence of the acceptor also enable to compute the FRET efficiency as $E_{FRET}=\Gamma_{FRET}/(\Gamma_{FRET}+\Gamma_{D})=1-\Gamma_{D}/\Gamma_{DA}$ (Fig.~\ref{FigTCSPC}f). As for the analysis based on the fluorescence bursts (Fig.~\ref{FigBursts}g,h), we observe that the FRET efficiencies deduced from the decay traces appear marginally affected by the nanoaperture in the case of short D-A separations.

\begin{figure}
\begin{center}
\includegraphics{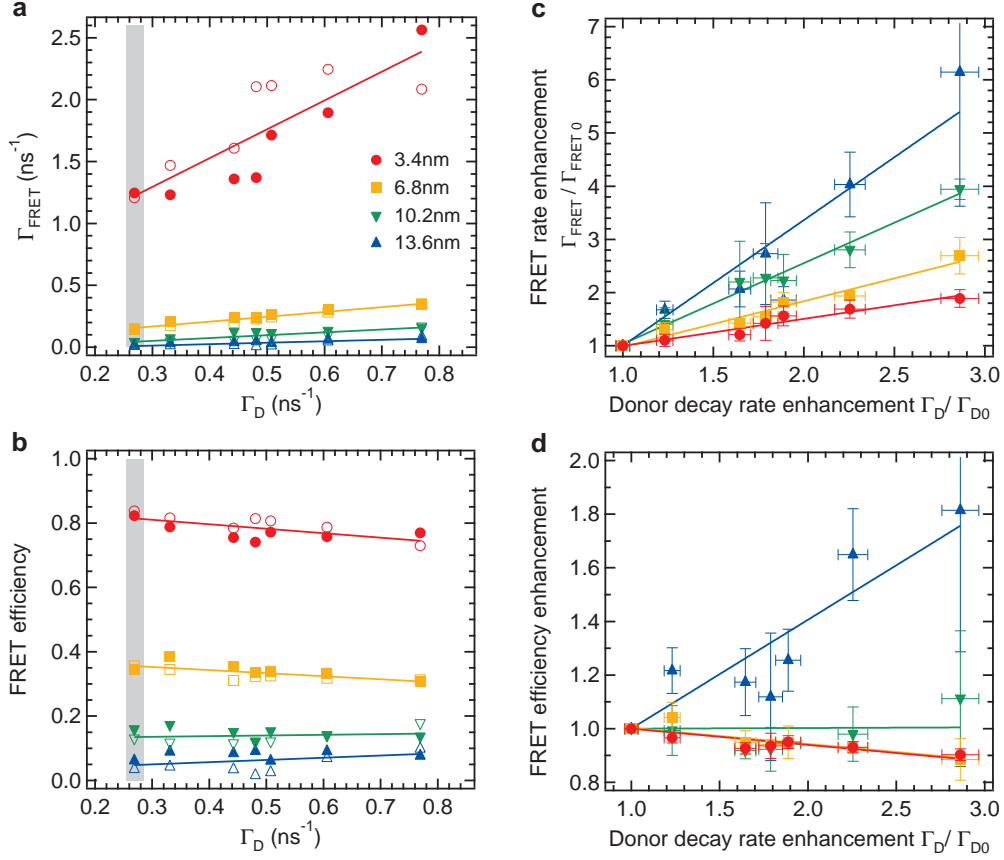}
\caption{\textbf{Nanophotonic control of FRET rate and efficiency.} (a) FRET rate $\Gamma_{FRET}$ and (b) FRET efficiency $E_{FRET}$ for different donor-acceptor separations as function of the donor-only decay rate $\Gamma_D$ that is proportional to the LDOS. Filled markers denote the data points deduced from the donor lifetime reduction, while empty markers denote data points deduced from fluorescence burst analysis. The lines are numerical fits of the average FRET rate between the two measurement methods, and show a linear dependence of the FRET rate with the LDOS for all donor-acceptor separations. The grey shaded region indicate the FRET rates obtained for the confocal reference. The evolution with the LDOS is further evidenced by normalizing the data in (a,b) by the values found for the confocal reference. This results in the graphs in (c,d) representing the normalized enhancement of the FRET rate and efficiency as function of the normalized enhancement of the donor-only decay rate. Horizontal error bars indicate one standard deviation of the measurements, vertical error bars are the difference between the two measurement methods for the average FRET rate and efficiency.} \label{FigLDOS}
\end{center}
\end{figure}

To establish the effect of the LDOS on the F\"{o}rster transfer, we display in Fig.~\ref{FigLDOS}a,b the FRET rate $\Gamma_{FRET}$ and efficiency $E_{FRET}$ as function of the donor-only decay rate $\Gamma_{D}$ which is proportional to the LDOS at the donor emission wavelength. A strength of our study is that we use two different approaches to quantify $\Gamma_{FRET}$ and $E_{FRET}$ based either on the donor lifetime reduction, or the acceptor fluorescence bursts (see Methods). The data points in Fig.~\ref{FigLDOS}a,b summarize the results: filled markers are deduced from the donor lifetime reduction and empty markers are deduced from fluorescence burst analysis. Both approaches converge to similar values, confirming the general trend. As our main result, our data clearly demonstrate a linear dependence of the FRET rate on the LDOS for all four D-A separations (Fig.~\ref{FigLDOS}a and Supplementary Information Fig.~9). This result is further confirmed by the slight variations of the FRET efficiency as the LDOS changes (Fig.~\ref{FigLDOS}b), which was already observed as function of the aperture diameter in Fig.~\ref{FigBursts}g,h and Fig.~\ref{FigTCSPC}f: to maintain a similar ratio $E_{FRET}=\Gamma_{FRET}/(\Gamma_{FRET}+\Gamma_{D})$, the FRET rate $\Gamma_{FRET}$ must increase accordingly as $\Gamma_{D}$ increases. Our results are in agreement with the ensemble-based experimental findings on microcavities \cite{Andrew00} and the dependence predicted theoretically \cite{Carminati,Enderlein}, and in contradiction with the conclusions found near a planar mirror \cite{Blum12}. We believe that the different D-A separations, the brighter emission rates and the larger LDOS variations in our case further ease assessing the LDOS influence on the FRET process.

To go further into the analysis, we display the FRET rate enhancement  $\Gamma_{FRET}/\Gamma_{FRET 0}$ as function of the LDOS enhancement $\Gamma_{D}/\Gamma_{D 0}$ by normalizing the results in Fig.~\ref{FigLDOS}a,b with the values found for the confocal reference ($\Gamma_{FRET 0}$, $\Gamma_{D 0}$). This representation reveals better the effect of the LDOS for the cases of large D-A separations and weak FRET rates and efficiencies. The global trend observed is a larger enhancement of the FRET rate as the D-A distance is increased: typically a 2x enhancement of the FRET rate is observed for a D-A distance of 3.4~nm when the LDOS is increased by $\sim3$x, whereas the FRET rate enhancement can be as large as 6x when the D-A distance is 13.6~nm. To explain the fact that the FRET rate enhancement and the LDOS enhancement do not necessarily coincide, it should be kept in mind that the LDOS enhancement (as we define it) amounts to the increase in the total decay rate and include both radiative and nonradiative (ohmic losses) energy decay channels, while we expect the FRET rate enhancement to be proportional to the radiative part of the LDOS enhancement \cite{Enderlein}.

\begin{figure}
\begin{center}
\includegraphics{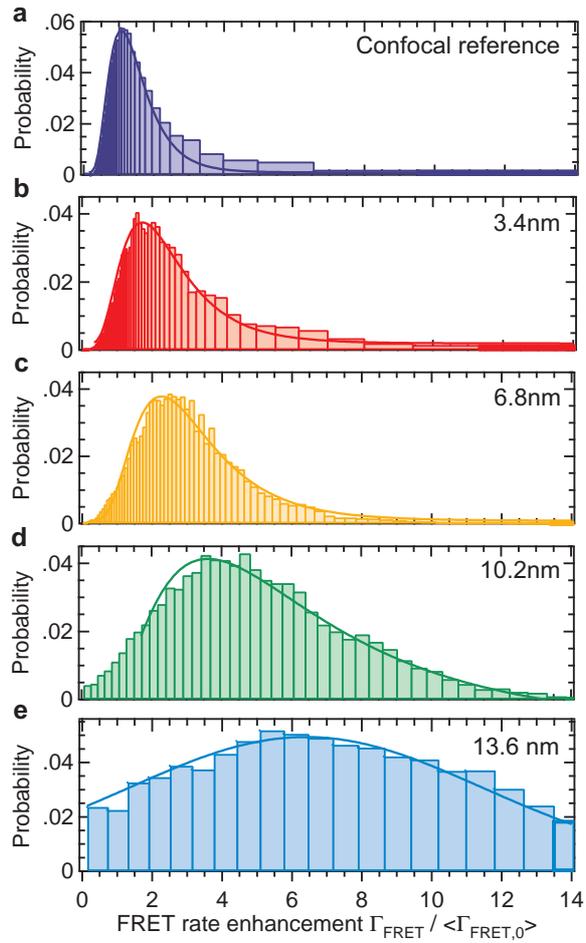}
\caption{\textbf{Distribution of FRET rates in nanoapertures} as compared to the average FRET rate found in the confocal reference. The histograms in (a) is the confocal reference for a 3.4~nm donor-acceptor separation, while data in (b-e) was taken in a 160~nm gold nanoaperture with increasing donor-acceptor distances. Lines are numerical fits with a log-normal distribution.} \label{FigHistog}
\end{center}
\end{figure}

Beyond the average values of FRET rates and efficiencies, monitoring the fluorescence bursts from single molecules allows to recover the statistical distribution of the FRET rates. This is obtained by reformulating the formula defining the FRET efficiency: $\Gamma_{FRET}= \Gamma_{D} \,\, E_{FRET}/(1-E_{FRET})$, and using the separate measurements of the FRET efficiency histograms (Fig.~\ref{FigBursts}h) and the donor-only decay rates $\Gamma_{D}$ (Fig.~\ref{FigTCSPC}b). Figure~\ref{FigHistog} depicts the distribution of the measured FRET rates enhancement $\Gamma_{FRET}/\langle \Gamma_{FRET 0}\rangle$, where $\langle \Gamma_{FRET 0}\rangle$ is the average FRET rate found for the confocal reference. As for Fig.~\ref{FigLDOS}c, the global trend is a larger enhancement of the FRET rate as the D-A distance increases. Furthermore, the statistical distributions show that significant FRET rate enhancement above 5-fold are reproducibly observed, demonstrating that nanophotonics can be used to enhance the near-field energy transfer.

In summary, we report enhanced energy transfer within single donor-acceptor fluorophore pairs confined in gold nanoapertures, and demonstrate experimentally that the F\"{o}rster energy transfer rate depends linearly on the LDOS. By increasing the donor dipole oscillator strength, the nanoaperture allows to efficiently transfer the energy to the acceptor dipole in the near-field. These findings are important as they unlock the potential application of the large nanophotonic toolbox for single emitter fluorescence control to further enhance the FRET process broadly used in single molecule spectroscopy applied to life sciences. An additional advantage is brought by the appealing property of nanoapertures to perform single molecule analysis at high physiological concentrations \cite{Levene03,TinnefeldRev13}, providing a new class of substrates for enhanced single molecule FRET analysis.

\section*{Methods}

\subsection*{Nanoaperture sample fabrication}
Nanoapertures are milled by focused ion beam (FEI Strata Dual Beam 235) on 150~nm thick gold films deposited using thermal evaporation on standard 150~$\mu$m thick glass
coverslips.

\subsection*{DNA synthesis and preparation}
Double-stranded DNA constructs of 51 base pairs length are designed with a one donor label Atto550 on the forward strand, and one acceptor label Atto647N on the reverse strand, at varying distances, so that the donor and acceptor are separated by 10, 20, 30 or 40 base pairs from each other to make at 3.4, 6.8, 10.2, or 13.6~nm varying distance respectively. As 10 base pairs make a complete turn on the DNA double strand, the choice of D-A separation as multiples of 10 base pairs avoids taking into account the complex three-dimensional structure of DNA to estimate the D-A distance \cite{Deniz99}. All constructs are purchased from IBA GmbH, Goettingen, Germany.

The forward strand sequence is

5'~CCTGAGCGTACTGCAGGATAGCCTATCGCGTGTCATATGCTGT$\mathrm{\mathbf{T_D}}$CAGTGCG~3'.

The reverse strand sequence is

5'~CGCACTGAACAGCATA$\mathrm{\mathbf{T_{10}}}$GACACGCGA$\mathrm{\mathbf{T_{20}}}$AGGCTATCC$\mathrm{\mathbf{T_{30}}}$GCAGTACGC$\mathrm{\mathbf{T_{40}}}$CAGG~3'.

The donor-only and acceptor-only references are constructed with the same sequences by replacing either the acceptor or the donor with unlabeled complementary strand respectively. The strands are annealed at 10~$\mu$M concentration in 20~mM Tris, 1~mM EDTA, 500~mM NaCl, 12~mM MgCl$_2$ buffer, and by heating to 95$^{\circ}$C for 5 min followed by slow cooling to room temperature. Samples were then stored at -28$^{\circ}$C. For single-molecule experiments labeled double stranded DNA stocks are diluted in a 10~mM Hepes-NaOH buffer, pH~7.5 (Sigma-Aldrich).

\subsection*{Experimental setup}
The experimental set-up is based on a confocal inverted microscope with a Zeiss C-Apochromat 63x 1.2NA water-immersion objective. The excitation source is a iChrome-TVIS laser (Toptica GmbH) delivering 3~ps pulses at 40~MHz repetition rate and 550~nm wavelength. Filtering the laser excitation is performed by a set of two bandpass filters (Chroma ET525/70M and Semrock FF01-550/88). The excitation power at the diffraction limited spot is set to 40~$\mu$W for all the experiments. Positioning the nanoaperture at the laser focus spot is obtained with a multi-axis piezoelectric stage (Polytech PI P-517.3CD). Dichroic mirrors (Chroma ZT594RDC and ZT633RDC) separate the donor and acceptor fluorescence light from the epi-reflected laser and elastically scattered light. The detection is performed by two avalanche photodiodes (Micro Photon Devices MPD-5CTC with 50~$\mu$m active surface and $<50$~ps timing jitter) with $600 \pm 20$~nm (Chroma ET605/70M and ET632/60M) and $670 \pm 20$~nm (Semrock FF01-676/37) fluorescence bandpass filters for the donor and acceptor channels respectively. The photodiode signal is recorded by a fast time-correlated single photon counting module (Hydraharp400, Picoquant GmbH) in time-tagged time-resolved (TTTR) mode. The temporal resolution of our setup for fluorescence lifetime measurements (width of the instrument response function) is 37~ps at half-maximum.

\subsection*{FRET analysis based on donor lifetime}
The decay traces obtained for the donor detection channel are fitted by a single exponential decay reconvoluted by the instrument response function (IRF) to extract the donor fluorescence lifetime using the commercial software Symphotime 64 (Picoquant GmbH). For the smallest aperture diameters, a background signal is detected due to the non-negligible photoluminescence from the gold film, and is accounted for in the lifetime analysis by adding a supplementary decay term with a fixed 5~ps characteristic time. For each aperture and each D-A separation, two sets of measurements are performed to determine the donor lifetime in the presence of the acceptor $\tau_{DA}=1/\Gamma_{DA}$ and the donor-only lifetime in the absence of acceptor $\tau_{D}=1/\Gamma_{D}$. The FRET efficiency is then obtained as $E_{FRET}=1-\Gamma_{D}/\Gamma_{DA}=1-\tau_{DA}/\tau_{D}$, and the FRET rate is obtained as $\Gamma_{FRET}=\Gamma_{DA}-\Gamma_{D}$.

\subsection*{FRET analysis based on acceptor fluorescence bursts}
For every detected fluorescence burst above the background noise, the number of detected photons in the acceptor channel $n_a$ and in the donor channel $n_d$ are recorded. Conceptually, these numbers are used to estimate the FRET efficiency as the ratio $n_a/(n_a+n_d)$ of acceptor emission events over all acceptor and donor events. Practically, several additional effects have to be taken into account to compensate for the direct excitation of the acceptor by the laser light, donor emission crosstalk into the acceptor channel, and differences in the quantum yields and detection efficiencies of the donor and acceptor emission. Using the commercial software Symphotime 64 (Picoquant GmbH), we apply the formula:
\begin{equation}
    E_{FRET} = \frac{n_a - \alpha n_d - n_{ao}^{de} }{n_a - \alpha n_d - n_{ao}^{de} + \gamma n_d}
\end{equation}
$\alpha$ is the crosstalk parameter defined as the ratio of donor-only fluorescence falling into the acceptor detection channel as compared to the donor-only signal detected in the donor channel. For all our measurements, $\alpha$ is fixed to a value of 0.17; we verified that $\alpha$ is not affected by the nanoaperture. $n_{ao}^{de}$ is the compensation parameter for the direct excitation of the acceptor dye by the laser light. This parameter was carefully measured for every nanoaperture by recording the average number of detected photons per burst when only the acceptor dye is present, and compensating for the slight differences of concentrations between the experiments (we use fluorescence correlation spectroscopy FCS to estimate the number of detected molecules and the molecular concentration in each experimental run, see Supplementary Information). Lastly, $\gamma=\eta_a \phi_a / \eta_d \phi_d$ accounts for the differences in quantum yields ($\phi_a$ and $\phi_d$) and fluorescence detection efficiencies ($\eta_a$ and $\eta_d$) between the acceptor and donor. For the confocal reference and the nanoapertures, we estimate $\gamma=1.3$ in the case of our setup. From the measurement of the FRET efficiency from fluorescence bursts, we deduce the FRET rate as $\Gamma_{FRET}= \Gamma_{D} E_{FRET}/(1-E_{FRET})$ where $\Gamma_{D}=1/\tau_{D}$ is the donor-only decay rate obtained from time-correlated lifetime measurements.

\subsection*{Supporting Information: Fluorescence correlation spectroscopy analysis}

Fluorescence correlation spectroscopy (FCS) is a versatile method to analyze the fluorescence time trace from a single molecule diffusing in solution, and quantify the number of molecules $N$ contributing to the detected fluorescence signal and their mean translational diffusion time $\tau_d$ \cite{Maiti,Fluobouquin}. FCS is based on the statistical analysis of the temporal fluctuations affecting the fluorescence intensity by computing the second order correlation of the fluorescence intensity time trace $g^{(2)}(\tau) = \langle F(t).F(t+\tau) \rangle / \langle F(t) \rangle ^2$, where $F(t)$ is the time-dependent fluorescence signal, $\tau$ the delay (lag) time, and $\langle . \rangle$ stands for time averaging. Analysis of the correlation function is based on a three dimensional Brownian diffusion model \cite{Maiti,Fluobouquin}:
\begin{equation}\label{Eq:diffFCS}
   g^{(2)}(\tau) = \frac{1}{N}\, \left( 1 - \frac{\langle B \rangle}{\langle F \rangle}\right)^2 \, \left[1 + n_T \exp \left(-\frac{\tau}{\tau_{b_T}} \right) \right]   \frac{1}{(1+\tau/\tau_d)\sqrt{1+s^2 \, \tau/\tau_d}}
\end{equation}
\noindent where $N$ is the total number of molecules, $\langle F \rangle$ the total signal, $\langle B \rangle$ the background noise, $n_T$ the amplitude of the dark state population, $\tau_{b_T}$ the dark state blinking time, $\tau_d$ the mean diffusion time and $s$ the ratio of transversal to axial dimensions of the analysis volume, which we set to $s=1$ for the nanoapertures and $s=0.2$ for the open solution reference.\cite{Davy08}. The background noise $\langle B \rangle$ originates mainly from the back-reflected laser light and from gold autofluorescence. At 20~$\mu$W excitation power, it is typically below 1~kHz, which appears negligible as compared to the fluorescence count rates per molecule in the nanoapertures. As a consequence of the stochastic nature of the FCS technique, all the presented fluorescence data are spatially averaged over all the possible molecule orientations and positions inside the detection volume.

Figure 6 presents typical FCS correlation data recorded in confocal diffraction-limited mode and in a 160~nm gold nanoaperture. The diffusion time reduction in the nanoaperture validates the absence of DNA binding to the gold surfaces in our case. The same correlation amplitude found in the donor and acceptor channel validates the fluorophore pair labeling of the DNA and the optical alignment.

\subsection*{Supporting Information: Fluorescence enhancement factors}

We normalize the average value of the fluorescence intensity $F$ by the average number of molecules $N$ (deduced by FCS) to compute the fluorescence count rate per molecule $CRM = F/N$. This information is then used to compare to the count rate per molecule found for the confocal reference, and estimate the fluorescence enhancement factor $\eta_F = CRM_{aper}/CRM_{sol}$ as function of the aperture diameter (Fig.~7). Comparable fluorescence enhancement factors are found for the donor and the acceptor emission. This demonstrates the broadband spectral response of the nanoapertures.

\subsection*{Supporting Information: Fluorescence decay traces and lifetimes}

Figure 8a-d display the time-correlated fluorescence decay traces for the different samples as function of the nanoaperture diameter. The fluorescence lifetimes extracted from the single exponential fits reconvoluted by the instrument response function are summarized in Fig. 8e,f. A similar lifetime reduction is obtained for the donor-only and the acceptor-only, showing that the nanoaperture response is broad spectrally and the LDOS is affected similarly at the donor emission and acceptor emission wavelengths. Moreover, a clear reduction of the donor fluorescence lifetime is observed for all nanoapertures as the donor-acceptor distance is reduced.

\subsection*{Supporting Information: FRET rates and efficiencies for large D-A separations}

At large donor-acceptor separations, the relatively weak FRET rates and efficiencies are difficult to visualize on the same vertical range as the results for the short 3.4~nm D-A separations (Fig.~4a,b of the main document). Figure~9 thus focus on the FRET rates and efficiencies for 10.2 and 13.6~nm D-A separations. The data is identical to Fig.~4a,b of the main document, only the vertical range is changed.


\section*{Acknowledgments}
We thank Thomas W. Ebbesen for stimulating discussions and support and Elo\"{\i}se Devaux for help with the FIB. The research leading to these results has received funding from the European Commission's Seventh Framework Programme (FP7-ICT-2011-7) under grant agreement ERC StG 278242 (ExtendFRET). PG is on leave from Institute for Space Sciences, Bucharest-M\u agurele RO-077125, Romania.


\newpage

\newpage

\newpage

\newpage

\newpage

\newpage

\begin{figure}[h!]
\begin{center}
\includegraphics{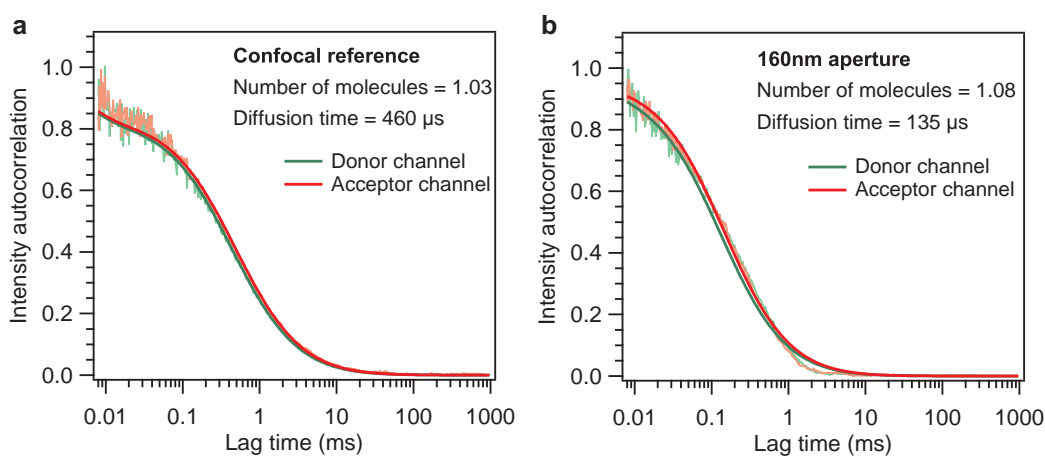}
\caption{\textbf{Supporting information: Fluorescence correlation spectroscopy validates the diffusion process and the number of molecules}. (a) Typical FCS correlation data in confocal mode and with a 160~nm nanoaperture (b) on donor-acceptor pairs with 6.8~nm separation. Thin lines are raw FCS data, thick lines are numerical fits assuming 3D Brownian diffusion. The concentration is set so as to have about one molecule in the detection volume in each case (confocal 1.7~nM, nanoaperture 0.5~$\mu$M).} \label{SIFigFCS}
\end{center}
\end{figure}

\newpage

\begin{figure}[h!]
\begin{center}
\includegraphics{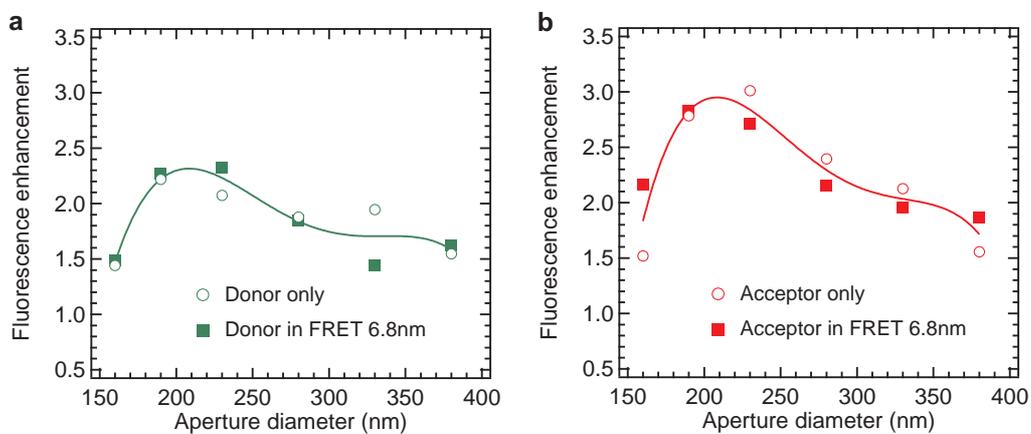}
\caption{\textbf{Supporting information: Nanoapertures enhance both donor and acceptor fluorescence}. Fluorescence enhancement factors respective to the confocal reference for (a) the donor fluorescence emission and (b) the acceptor fluorescence emission. Fluorescence enhancement factors are deduced from the brightness per molecule measured by FCS, and always refer to the same sample (with or without acceptor) measured in the confocal condition. The presence of the complementary fluorophore to form the FRET pair is not found to affect the fluorescence enhancement factor in the nanoaperture. Lines are guide to the eyes.} \label{SIFigEnh}
\end{center}
\end{figure}

\newpage

\begin{figure}
\begin{center}
\includegraphics{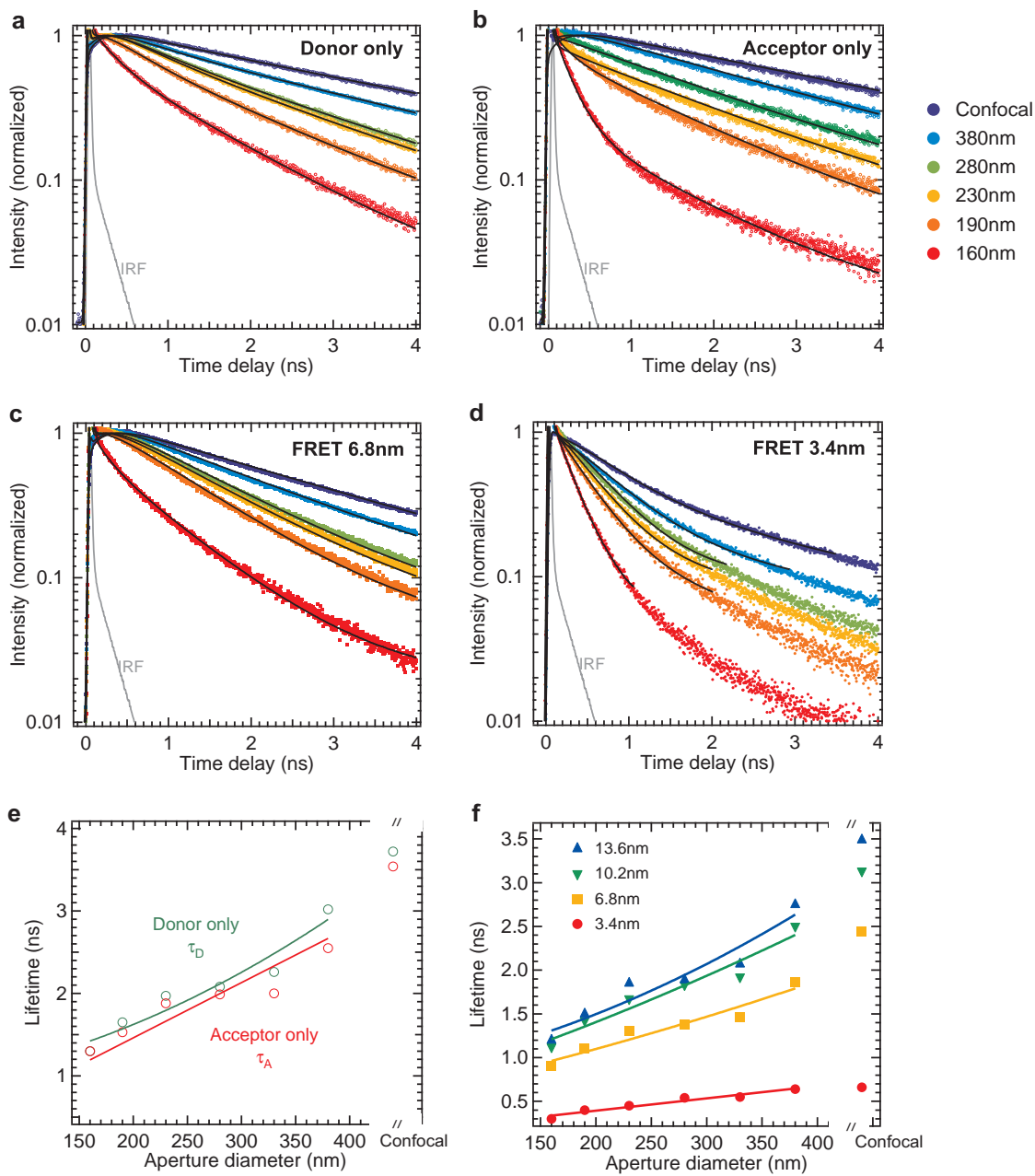}
\caption{\textbf{Supporting information: Fluorescence photodynamics in nanoapertures}. (a) Normalized donor decay traces when no acceptor is present, (b) normalized acceptor decay traces when no donor is present, (c) normalized donor decay traces when the acceptor separated by 6.8~nm, and (d) normalized donor decay traces when the acceptor separated by 3.4~nm. The black lines are the single-exponential numerical fits convoluted by the instrument response function (IRF). From top to bottom, the different colors are associated with the confocal case or with nanoapertures of decreasing diameters from 380 to 160~nm. (e) Donor-only and acceptor-only fluorescence lifetime as function of the nanoaperture diameter, obtained from the data in (a,b). (f) Donor fluorescence lifetime in the presence of the acceptor as function of the nanoaperture diameter.} \label{SIFigTCSPC}
\end{center}
\end{figure}

\newpage

\begin{figure}[h!]
\begin{center}
\includegraphics{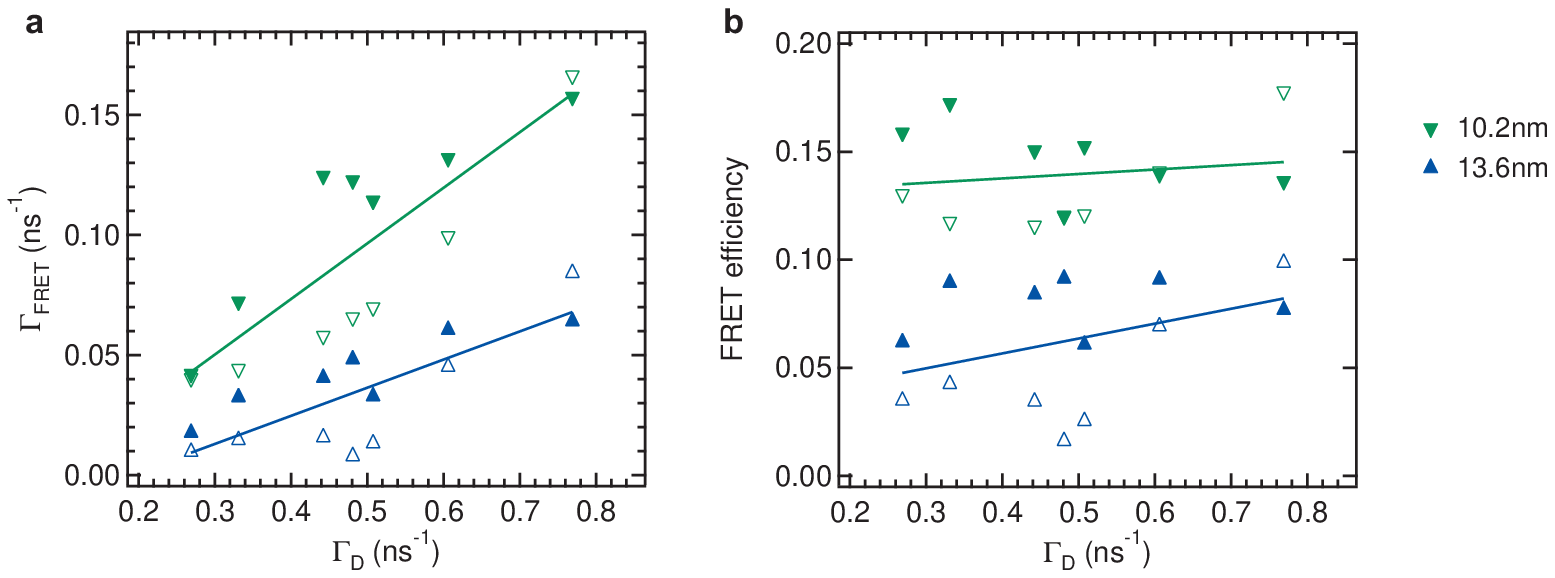}
\caption{\textbf{Supporting information: A linear dependence of the FRET rate on the LDOS is found also for large donor-acceptor separations}. This figure displays the same data as in Fig.~4 (a,b) with a smaller vertical range to better visualize the LDOS influence in the cases of donor-acceptor separations larger than 10~nm. (a) FRET rate $\Gamma_{FRET}$ and FRET efficiency $E_{FRET}$ (b) for different donor-acceptor separations as function of the donor-only decay rate $\Gamma_D$ that is proportional to the LDOS. Filled markers denote the data points deduced from the donor lifetime reduction, while empty markers denote data points deduced from fluorescence burst analysis. The lines are numerical fits of the average FRET rate between the two measurement methods, and show a linear dependence with the LDOS.} \label{SIFigZoom}
\end{center}
\end{figure}


\begin{thebibliography}{99}

\bibitem{Kuhlbrandt91} K\"{u}hlbrandt W, \& Wang DN. Three-dimensional structure of plant light-harvesting complex determined by electron crystallography. \textit{Nature} \textbf{350}, 130-134 (1991).

\bibitem{Grondelle94} van Grondelle, R., Dekker, J. P., Gillbro, T., \& Sundstr\"{o}m, V. Energy transfer and trapping in photosynthesis. \textit{Biochim. Biophys. Acta Bioenerg.} \textbf{1187}, 1-65 (1994).

\bibitem{Niek13} Hildner R, Brinks D, Nieder JB, Cogdell RJ \& van Hulst NF. Quantum coherent energy transfer over varying pathways in single light-harvesting complexes. \textit{Science} \textbf{340}, 1448-1451 (2013).

\bibitem{Farrell09} Farrell, D.J. \& Ekins-Daukes N.J. Photovoltaic technology: Relay dye boosts efficiency. \textit{Nat. Photon.} \textbf{3}, 373-374 (2009).

\bibitem{Shankar09} Shankar, K., Feng, X. \& Grimes, C. A. Enhanced harvesting of red photons in nanowire solar cells: Evidence of resonance energy transfer. \textit{ACS Nano} \textbf{3}, 788-794 (2009).

\bibitem{Baldo00} Baldo, M.A., Thompson, M.E. \& Forrest S.R. High-efficiency fluorescent organic light-emitting devices using a phosphorescent sensitizer. \textit{Nature} \textbf{403}, 750-753 (2000).

\bibitem{Medintz03} Medintz, I. L., Clapp, A. R., Mattoussi, H., Goldman, E. R., Fisher, B., \& Mauro, J. M.. Self-assembled nanoscale biosensors based on quantum dot FRET donors. \textit{Nat. Mater.} \textbf{2}, 630-638 (2003).

\bibitem{Forster48} F\"{o}rster, T. Zwischenmolekulare energiewanderung und fluoreszenz. \textit{Ann. Phys.} \textbf{437}, 55-75 (1948).

\bibitem{Roy08} Roy, R., Hohng, S. \& Ha, T. A practical guide to single-molecule FRET. \textit{Nat. Methods} \textbf{5}, 507-516 (2008).

\bibitem{Weiss00} Weiss S. Measuring conformational dynamics of biomolecules by single molecule fluorescence spectroscopy. \textit{Nat. Struct. Biol.}\textbf{7}, 724-729 (2000).

\bibitem{Schuler08} Schuler, B. \& Eaton, W.A. Protein folding studied by single-molecule FRET. \textit{Curr. Opin. Struct. Biol.} \textbf{18}, 16-26 (2008).

\bibitem{Purcell46} Purcell, E. M. Spontaneous emission probabilities at radio frequencies. \textit{Phys. Rev.} \textbf{69}, 681 (1946).

\bibitem{Drexhage70} Drexhage, K. H. Influence of a dielectric interface on fluorescence decay time. \textit{J. Lumin.} \textbf{1}, 693-701 (1970).

\bibitem{Novotnybook} Novotny, L. \& Hecht, B. Principles of Nano-Optics. Cambridge University Press, Cambridge (2006).

\bibitem{Barnes98} Barnes, W. L. Fluorescence near interfaces: the role of photonic mode density. \textit{J. Mod. Opt.} \textbf{45}, 661-699 (1998).

\bibitem{Haroche} Goy, P., Raimond, J.M., Gross, M. \& Haroche, S. Observation of Cavity-Enhanced Single-Atom Spontaneous Emission. \textit{Phys. Rev. Lett.} \textbf{50}, 1903-1906 (1983).

\bibitem{Yablo} Yablonovitch, E. Inhibited spontaneous emission in solid-state physics and electronics. \textit{Phys. Rev. Lett.} \textbf{58}, 2059-2062 (1987).

\bibitem{Lodahl04} Lodahl, P., Van Driel, A. F., Nikolaev, I. S., Irman, A., Overgaag, K., Vanmaekelbergh, D., \& Vos, W. L. Controlling the dynamics of spontaneous emission from quantum dots by photonic crystals. \textit{Nature} \textbf{430}, 654-657 (2004).

\bibitem{NovotnyRev11} Novotny, L. \& van Hulst, N. Antennas for light. \textit{Nature Photon.} \textbf{5}, 83-90 (2011).

\bibitem{Moerner09} Kinkhabwala, A., Yu, Z. F., Fan, S. H., Avlasevich, Y., Mullen, K. \& Moerner, W. E. Large single-molecule fluorescence enhancements produced by a bowtie nanoantenna. \textit{Nature Photon.} \textbf{3}, 654-657 (2009).

\bibitem{Hopmeier99} Hopmeier,  M., Guss, W., Deussen, M., G\"{o}bel, E.O. \& Mahrt, R.F. Enhanced dipole-dipole interaction in a polymer microcavity. \textit{Phys. Rev. Lett.} \textbf{82}, 4118-4121 (1999).

\bibitem{Andrew00} Andrew, P. \& Barnes, W.L. F\"{o}rster energy transfer in an optical microcavity. \textit{Science} \textbf{290}, 785-788 (2000).

\bibitem{Finlayson01} Finlayson, C. E., Ginger, D. S. \& Greenham, N. C. Enhanced F\"{o}rster energy transfer in organic/inorganic bilayer optical microcavities. \textit{Chem. Phys. Lett.} \textbf{338}, 83-87 (2001).

\bibitem{Nakamura05} Nakamura, T., Fujii, M., Imakita, K. \& Hayashi, S.  Modification of energy transfer from Si nanocrystals to Er$^{3+}$ near a Au thin film. \textit{Phys. Rev. B} \textbf{72}, 235412 (2005).

\bibitem{Dung} Dung, H. T., Kn\"{o}ll, L. \& Welsch, D. G.  Intermolecular energy transfer in the presence of dispersing and absorbing media. \textit{Phys. Rev. A} \textbf{65}, 043813 (2002).

\bibitem{Colas} Colas des Francs, G., Girard, C. \& Martin, O. J.  Fluorescence resonant energy transfer in the optical near field. \textit{Phys. Rev. A} \textbf{67}, 053805 (2003).

\bibitem{Govorov07} Govorov, A. O., Lee, J. \& Kotov, N. A.  Theory of plasmon-enhanced F\"{o}rster energy transfer in optically excited semiconductor and metal nanoparticles. \textit{Phys. Rev. B} \textbf{76}, 125308 (2007).

\bibitem{Carminati} Vincent, R. \& Carminati, R.  Magneto-optical control of F\"{o}rster energy transfer. \textit{Phys. Rev. B} \textbf{83}, 165426 (2011).

\bibitem{Enderlein} Enderlein, J. Modification of F\"{o}rster resonance energy transfer efficiency at interface. \textit{Int. J. Mol. Sci.} \textbf{13}, 15227-15240 (2012).

\bibitem{Nakamura06} Nakamura, T., Fujii, M., Miura, S., Inui, M. \& Hayashi, S.  Enhancement and suppression of energy transfer from Si nanocrystals to Er ions through a control of the photonic mode density. \textit{Phys. Rev. B} \textbf{74}, 045302 (2006).

\bibitem{Polman05} de Dood, M. J. A., Knoester, J., Tip, A. \& Polman, A.  F\"{o}rster transfer and the local optical density of states in erbium-doped silica. \textit{Phys. Rev. B} \textbf{71}, 115102 (2005).

\bibitem{Blum12} Blum, C., Zijlstra, N., Lagendijk, A., Wubs, M., Mosk, A. P., Subramaniam, V. \& Vos, W. L.  Nanophotonic control of the F\"{o}rster resonance energy transfer efficiency. \textit{Phys. Rev. Lett.} \textbf{109}, 203601 (2012).

\bibitem{Lakowicz07} Zhang, J., Fu, Y., Chowdhury, M. H., \& Lakowicz, J. R. Enhanced F\"{o}rster resonance energy transfer on single metal particle. 2. dependence on donor-acceptor separation distance, particle size, and distance from metal surface. \textit{J. Phys. Chem. C} \textbf{111}, 11784-11792 (2007).

\bibitem{Fore07} Fore, S., Yuen, Y., Hesselink, L., \& Huser, T. Pulsed-interleaved excitation FRET measurements on single duplex DNA molecules inside C-shaped nanoapertures. \textit{Nano Lett.} \textbf{7}, 1749-1756 (2007).

\bibitem{Kolaric07} Kolaric, B., Baert, K., Van der Auweraer, M., Vall\'{e}e, R. A., and, \& Clays, K. Controlling the fluorescence resonant energy transfer by photonic crystal band gap engineering. \textit{Chem. Mater.} \textbf{19}, 5547-5552 (2007).

\bibitem{Hohenester08} Reil, F., Hohenester, U., Krenn, J. R., \& Leitner, A. F\"{o}rster-type resonant energy transfer influenced by metal nanoparticles. \textit{Nano Lett.} \textbf{8}, 4128-4133 (2008).

\bibitem{Yang08} Yang, Z., Zhou, X., Huang, X., Zhou, J., Yang, G., Xie, Q.,Sun, L. \& Li, B. Energy transfer between fluorescent dyes in photonic crystals. \textit{Opt. Lett.} \textbf{33}, 1963-1965 (2008).

\bibitem{Bradley08} Komarala, V. K., Bradley, A. L., Rakovich, Y. P., Byrne, S. J., Gunko, Y. K., \& Rogach, A. L. Surface plasmon enhanced F\"{o}rster resonance energy transfer between the CdTe quantum dots. \textit{Appl. Phys. Lett.} \textbf{93}, 123102-123102 (2008).

\bibitem{Bradley11} Lunz, M., Gerard, V. A., Gun'ko, Y. K., Lesnyak, V., Gaponik, N., Susha, A. S., Rogach, A.L. \& Bradley, A. L. Surface plasmon enhanced energy transfer between donor and acceptor CdTe nanocrystal quantum dot monolayers. \textit{Nano Lett.} \textbf{11}, 3341-3345 (2011).

\bibitem{Bradley12} Zhang, X., Marocico, C. A., Lunz, M., Gerard, V. A., Gun'ko, Y. K., Lesnyak, V., Gaponik, N., Susha, A.S., Rogach, A.L. \& Bradley, A. L. Wavelength, Concentration, and distance dependence of nonradiative energy transfer to a plane of gold nanoparticles. \textit{ACS Nano} \textbf{6}, 9283-9290  (2012).

\bibitem{Rigneault05} Rigneault, H., Capoulade, J., Dintinger, J., Wenger, J., Bonod, N., Popov, E., Ebbesen, T. W. \& Lenne, P. F. Enhancement of single-molecule fluorescence detection in subwavelength apertures. \textit{Phys. Rev. Lett.} \textbf{95}, 117401 (2005).

\bibitem{Wenger08} Wenger, J., G\'{e}rard, D., Bonod, N., Popov, E., Rigneault, H., Dintinger, J., Mahboub, O. \& Ebbesen, T.W. Emission and excitation contributions to enhanced single molecule fluorescence by gold nanometric apertures. \textit{Opt. Express} \textbf{16}, 3008-3020 (2008).

\bibitem{LutzACS13} Langguth, L., Punj, D., Wenger, J. \& Koenderink, A.F. Plasmonic Band Structure Controls Single-Molecule Fluorescence. \textit{ACS Nano} \textbf{7}, 8840-8848 (2013).



\bibitem{Deniz99} Deniz, A.A., Dahan, M., Grunwell, J.R., Ha, T., Faulhaber, A.E., Chemla, D.S., Weiss, S. \& Schultz P.G. Single-pair fluorescence resonance energy transfer on freely diffusing molecules: observation of F\"{o}rster distance dependence and subpopulations. \textit{Proc. Natl. Acad. Sci. USA} \textbf{96}, 3670-3675 (1999).

%
%
%

\bibitem{Levene03} Levene, M. J., Korlach, J., Turner, S. W., Foquet, M., Craighead, H. G. \& Webb, W. W. Zero-mode waveguides for single-molecule analysis at high concentrations. \textit{Science} \textbf{299}, 682-686 (2003).

\bibitem{TinnefeldRev13} Holzmeister, P., Acuna, G.P., Grohmann, D. \& Tinnefeld, P. Breaking the concentration limit of optical single-molecule detection. \textit{Chem. Soc. Rev.}, DOI: 10.1039/c3cs60207a


\bibitem{Maiti} Maiti, S., Haupts, U. \& Webb, W.W. Fluorescence correlation spectroscopy: diagnostics for sparse molecules. \textit{Proc. Nat. Acad. Sci. USA} \textbf{94}, 11753-11757 (1997).

\bibitem{Fluobouquin} Zander, C., Enderlein J. \& Keller, R. A. Single-Molecule Detection in Solution - Methods and Applications, VCH-Wiley, Berlin/New York, 2002.

\bibitem{Davy08} G\'{e}rard, D.; Wenger, J.; Bonod, N.; Popov, E.; Rigneault, H.; Mahdavi, F.; Blair, S.; Dintinger, J. \& Ebbesen, T.W. Nanoaperture-enhanced fluorescence: towards higher detection rates with plasmonic metals. \textit{Phys. Rev. B} \textbf{77}, 045413 (2008).

\end{thebibliography}
\end{document}